\documentclass[11pt]{article}
\usepackage{color}   %May be necessary if you want to color links
\usepackage{hyperref}
\hypersetup{linktocpage}
\usepackage[textwidth=15.2cm,textheight=22cm]{geometry}
\usepackage{amsmath,amssymb}
\usepackage{latexsym}
\usepackage{multicol}
\usepackage{graphicx}
\usepackage{caption}
\usepackage{cancel}
\usepackage{bbm}
\usepackage{bm}
\usepackage{cite}
\numberwithin{equation}{section}
\pdfoutput=1
\usepackage[T1]{fontenc}
\usepackage[latin9]{inputenc}
\usepackage[active]{srcltx}
\usepackage{esint}
\usepackage{cancel}
\usepackage{color}
\usepackage{ulem}
\usepackage{caption}
\allowdisplaybreaks[1]

\newcommand{\del}{\partial}
\newcommand{\be}{\begin{equation}}
\newcommand{\ee}{\end{equation}}
\newcommand{\bea}{\begin{eqnarray}}
\newcommand{\eea}{\end{eqnarray}}

\newcommand{\nn}{\nonumber}

\newcommand{\ie}{{\it i.e.\ }}

\newcommand{\ndt}{\noindent}

\definecolor{ggreen}{rgb}{0.0, 0.5, 0.20}

\begin{document}

	\begin{titlepage}
		\begin{flushright}    
			{\small $\,$}
		\end{flushright}
		\vskip 1cm
		\centerline{\Large{\bf{Residual gauge symmetry in light-cone electromagnetism}}}
		\vskip 1.5cm
		\centerline{\large{Sucheta Majumdar\footnote{sucheta.majumdar@ens-lyon.fr}}}
		\vskip .7cm      
	\centerline{\it{Univ Lyon, ENS de Lyon, CNRS, Laboratoire de physique,
}}
	\centerline{\it{UMR 5672, F-69342 Lyon, France}}
	\vskip 1.5cm
		\centerline{\bf {Abstract}}
		\vskip .5cm
		\ndt We analyze the residual gauge freedom in light-cone electromagnetism in four dimensions. The standard boundary conditions involved in the so-called \textit{$lc_2$ formalism}, which contains only the two physical degrees of freedom,  allow for a subset of residual gauge transformations. We relax the boundary conditions imposed on the fields in order to obtain all the residual gauge transformations. We compute the canonical generators for Poincar\'e and gauge transformations with these relaxed boundary conditions. This enables us to distinguish between the trivial (proper) and large (improper) gauge transformations in light-cone electromagnetism. We then employ the Newman-Penrose formalism to identify the incoming and outgoing radiation fields. We comment on the quadratic form structure of light-cone Hamiltonians, often encountered in $lc_2$ gauge theories.

		\vfill
	\end{titlepage}

\tableofcontents
\section{Introduction}
The light-cone quantization of gauge theories has been a subject of great importance in quantum field theory and string theory (see~\cite{Leibbrandt:1987qv, Brodsky:1997de,Strings} and references therein).  Based on Dirac's front form of relativistic dynamics~\cite{Dirac:1949cp}, the light-cone formulation uses one of the null coordinates $x^{\pm}$ as the time or evolution parameter. This usually makes the constraint equations solvable, allowing us to eliminate unphysical degrees of freedom from the theory, a feature that has proved to be fruitful in many instances. Due to the absence of gauge degrees of freedom, there is no need to introduce auxiliary fields or Fadeev-Popov ghosts while applying quantization techniques. The light-cone formalism for massless fields involving the two physical states (in four dimensions), known as the $lc_2$ formalism, reduces the complexity of the theories tremendously. Some noteworthy examples include loop computations in QCD~\cite{Brodsky:1997de}  and the proof of ultraviolet finiteness of the $\mathcal N=4$ super Yang- Mills theory~\cite{Brink:1982wv,Mandelstam:1982cb}.  The light-cone superspace approach simplifies supersymmetric theories to a great extent, often just in terms of a single complex scalar superfield~\cite{Bengtsson:1983pg}.
\par
In the light-cone formalism, even though the gauge freedom is said to be completely fixed, there remains some residual gauge symmetry. Residual gauge symmetry in a theory is often linked to large gauge transformations that do not vanish at asymptotic infinity. This begs the question $-$ what are the residual gauge transformations in the $lc_2$ formalism and how are these related to large gauge transformations? Recently, we had investigated this issue in light-cone gravity~\cite{Ananth:2020ojp,Ananth:2020ngt}, where the residual diffeomorphisms were shown to contain the BMS supertranslations that arise as an asymptotic symmetry at spatial or null infinity. 
\par 
Asymptotic symmetries have garnered much attention in the recent years following some exciting developments linking them to soft theorems for scattering amplitudes, Ward identities and the gravitational S-matrix.  Scattering amplitudes are best understood in on-shell frameworks that involve working in a physical gauge, such as the light-cone gauge . As a matter of fact, the light-front serves as a natural basis for studying scattering amplitudes. The dynamics in light-cone field theories occur on the light front, where the scattering of massless particles take place. In the light-cone formalism, many interesting amplitude structures, such as KLT relations and MHV amplitudes, appear at the level of the action~\cite{Gorsky:2005sf,Mansfield:2005yd,Ananth:2007zy,Ananth:2011hu}. Massless fields in $lc_2$ theories are described in a basis of their two helicity eigenstates, that appear in the scattering amplitudes. Thus, the light-cone analysis of residual gauge symmetries may offer a unique framework for exploring the links between scattering amplitudes and asymptotic symmetries.
\par
In this paper, we analyze the residual gauge freedom in light-cone electromagnetism in four dimensions. In Section 2, we review $lc_2$ electromagnetism and show that the standard boundary conditions first introduced by Kogut and Soper~\cite{Kogut:1969xa} only allow for a subset of residual gauge transformations. In the next section, we relax the boundary conditions imposed on the fields in order to obtain all the residual gauge transformations. We compute the canonical generators for Poincar\'e and gauge transformations with these relaxed boundary conditions. This enables us to distinguish between the trivial (proper) and large (improper) gauge transformations in light-cone electromagnetism. The trivial gauge transformations do not change the physical state of the system and hence, have zero charge in the canonical generator for gauge transformations. One can freely perform trivial gauge transformation without affecting the physical configuration of the fields. Improper gauge transformations have non-vanishing gauge charge and change the physical state of the system. In order to show that the light-cone Maxwell fields indeed correspond to electromagnetic radiation, we employ the Newman-Penrose formalism~\cite{Newman:1961qr} in Section 4 to identify the incoming and outgoing radiation fields. Surprisingly, we also rediscover the quadratic form structure of the light-cone Hamiltonian for electromagnetism. We comment on the possibility that there might be some hidden relations between the Newman-Penrose formalism and the quadratic form structure of Hamiltonians, often encountered in $lc_2$ gauge theories.

\section{Electromagnetism in $\textit{lc}_2$ formalism} 

%***************************************************************************************************************************************
\subsection{Light-cone electromagnetism}
In this section, we review the salient features of electromagnetism in the $lc_2$ formalism~\cite{Kogut:1969xa} and discuss the residual gauge freedom of the theory. We use the light-cone coordinates defined as
\be
 x^+ = \frac{x^0 +x^3}{\sqrt 2}\,, \quad x^- = \frac{x^0-x^3}{\sqrt 2}\, \quad x= \frac{x^1 + ix^2}{\sqrt 2}\, , \quad \bar{x} = \frac{x^1-ix^2}{\sqrt 2}  \,.
\ee
We relabel the light-cone coordinates for our convenience as $x^\mu = (s,\rho, x^I)$
\be
s =x^+\,, \quad  \rho = x^- \, ,\quad x^I \equiv (x, \bar x) \,.
\ee
The coordinate $s=x^+$ is considered to be the time or evolution parameter, such that the corresponding momentum $P_s = -P^\rho$ gives the Hamiltonian $H$. The initial data is specified on a constant $s$ null plane and the Hamiltonian $H$ governs the time evolution of the system off of this null plane, as shown in Fig.~(\ref{fig:null}). The details of the light-cone coordinates used in this paper can be found in Appendix~(\ref{appendix:A}). 
\begin{figure}
\centering
\includegraphics[width=10cm]{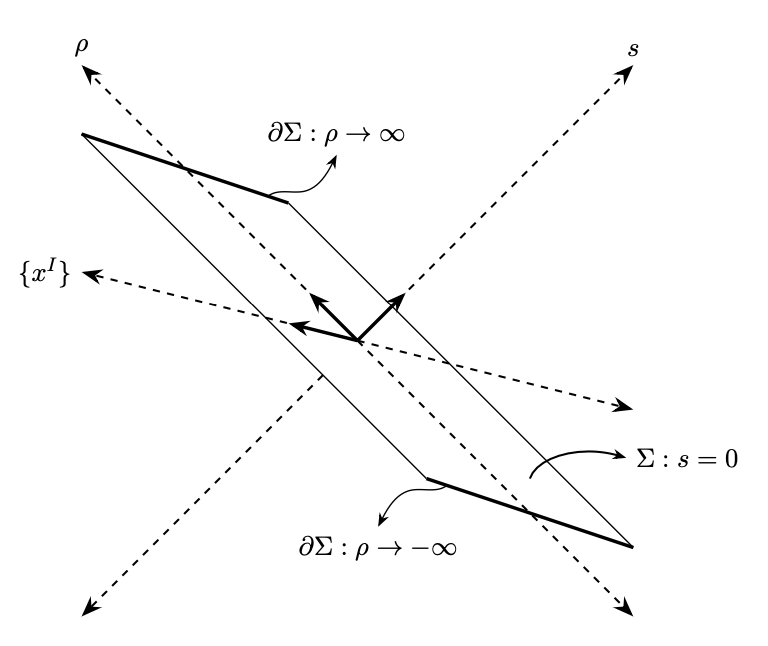}
\caption{The $s=0$ null plane in light-cone coordinates $(s, \rho, x^I)$} \label{fig:null}
\end{figure}
\par
We begin with the covariant action for electromagnetism in four dimensions
\be
\mathcal S =- \frac{1}{4}\int d^4x\, F^{\mu \nu} F_{\mu \nu} \,,
\ee	
with the field strength 
\be
F_{\mu \nu} = \del_\mu A_\nu - \del_\nu A_\mu\,.
\ee
\par
The light-cone gauge choice is to set $A^s (\mathbf{x})= - A_\rho (\mathbf{x})$ to zero\footnote{We shall often use $\mathbf{x}$ as a shorthand for $x^\mu$ to denote that an object depends on all four coordinates.}. As a consequence, the Maxwell's equations in light-cone coordinates split into three kinds:

\textit{i)} Constraint equation that does not involve time derivatives
\be \label{constraint0}
\del_\mu F^{\mu s} = 0\,,
\ee

\textit{ii)} Dynamical equation that governs the dynamics of the system
\be \label{dynamical0}
\del_\mu F^{\mu I} = 0\,,
\ee

\textit{iii)} Trivial equation that is identically satisfied if the above two equations hold
\be \label{trivial0}
\del_\mu F^{\mu \rho}= 0\,.
\ee

 We first focus on the constraint equation which yields
\bea \label{constraint}
\del_\rho^2 A^\rho + \del_\rho \del_I A^I ~=~ 0 \quad \Rightarrow \quad   \del_\rho A^\rho + \del_I A^I ~=~ \alpha(s, x^I)\,. \label{constraint-alpha}
\eea
As it does not contain any time derivatives, $\del_s$, it can be solved in order to eliminate the $A^\rho$ in terms of $A^I$
\be \label{Aminus}
A^\rho (\mathbf{x}) = \,  -\frac{\del^I}{\del_\rho} A_I (\mathbf{x})\, +\alpha(s, x^I) \, \rho + \beta(s, x^I) \,,
\ee
where the operator $\frac{1}{\del_\rho}$ represents an integration over $\rho$ and the functions $\alpha, \beta$ are the constants of integration associated with it. Formally, the $\frac{1}{\del_\rho}$ operator is defined using the Heaviside step function as~\cite{Kogut:1969xa}
\bea \label{inv deriv}
\frac{1}{\del_\rho} f(\mathbf{x}) &=& \frac{1}{2} \int d \rho' \, \epsilon(\rho - \rho') f(s, x^I, \rho')\,, \nn\\
\epsilon(y)&=& \Big\{\begin{array}{cc} \,
1 & y>0 \\
-1 & y<0\,.
\end{array}\,
\eea
 In the light-cone formulation, it is customary to set these integration constants to zero, such that $A^\rho$ is completely determined by the two remaining components $A^I$. This choice is often referred to as the $lc_2$ formalism as the theory now involves only the \textit{two} physical degrees of freedom. In this section, we will set these constants to zero, as we are interested in the $lc_2$ formalism. The case where these constants are non-vanishing will be discussed in the next section.
 \vskip 0.2cm
\ndt We now turn to the dynamical equation
\be \label{Dyn}
\del_\mu F^{\mu A} = (-2 \del_s \del_\rho + \triangle) A^I = \Box A^I (\mathbf{x}) =0 \,,
 \ee
with $\triangle$ being the two-dimensional Laplacian $\del^I\del_I$. Here, we have used the solution for $A_\rho$ in terms of $A^I$ from~\eqref{Aminus}. $A^I(\mathbf{x})$ are, therefore, the two propagating degrees of freedom that satisfy the wave equation. The trivial equation~\eqref{trivial} holds as an identity when the constraint and dynamical equations are satisfied.

%**************************************************************************************************************************************

\subsection{Kogut-Soper boundary conditions}

\ndt At this point, it is important to specify the boundary conditions imposed on the fields as we approach the limit $\rho \rightarrow \pm \infty$. In the light-cone gauge, the standard boundary conditions on the field strength, first introduced by Kogut and Soper~\cite{Kogut:1969xa}, are as follows. 
\be
F^{s\mu}\sim \mathcal O \left( \frac{1}{\rho^2}\right) \quad \text{as} \quad \rho \rightarrow \pm \infty \,.
\ee
On the vector fields, this implies 
\bea
A^\mu (\mathbf{x}) \sim \mathcal O \left( \frac{1}{\rho}\right)\,, \quad A^\mu (s, x^I, \rho = +\infty) = - A^\mu (s, x^I, \rho = -\infty) \,.
\eea
The second condition follows from symmetry arguments to ensure that the Hilbert space contains finite energy states~\cite{Steinhardt:1979it}. With these boundary conditions, the operation \eqref{inv deriv} is well-defined as $\rho$ tends to infinity. We can, thus, freely perform integrations by part over $\rho$ without worrying about surface terms.
\par
The time component of the field strength, $F^{\mu s}$ can be thought of as the light-cone electric field $\vec{E}$ with components $(F^{\rho s}, F^{As})$. Thus, the constraint equation may be considered as the light-cone analogue of the Gauss law $\vec{\nabla} . \vec{E} =0 $, which is consistent with the Kogut-Soper (KS) boundary conditions.
\par
We shall, therefore, assume that the fields satisfy the KS boundary conditions and admit the following asymptotic expansion in $\rho^{-1}$ 
\bea
A^I (\mathbf{x}) &=& \frac{A^I_{(0)} (s, x^J)}{\rho} + \frac{A^I_{(1)} (s, x^J)}{\rho^2} + \ldots \,,\\
A^\rho (\mathbf{x}) &=& \frac{A^\rho_{(0)}(s, x^J)}{\rho} + \frac{A^\rho_{(1)} (s, x^J)}{\rho^2} + \ldots \,.
\eea
From the constraint~\eqref{constraint}, we find at the leading order,
\be \label{boost-inv}
\del_I A^I_{(0)} (s, x^J)= 0 \,,
\ee 
and at the subleading orders,
\be \label{constraint2}
 A^\rho_{(n-1)} (s, x^J)\, = \frac{1}{(n+1)}\del_I A^I_{(n)} (s, x^J)\,,  \quad n = 1,2,...
\ee
So, $A^\rho$ is completely specified by $A^I$ at each order in the expansion. As we will discuss later, the condition~\eqref{boost-inv} is crucial for well-defined canonical generators for Lorentz boosts.  This completes the solution space of electromagnetism in the light-cone gauge with the KS boundary conditions.

%***************************************************************************************************************************************
\subsection{Canonical generators and light-cone Poincar\'e algebra}
\label{section:2.3}
\ndt
\ndt
The light-cone action for electromagnetism in terms of $A^I$ reads
\be
\mathcal{S}[A^I] = \frac{1}{2}  \int d^4 \mathbf{x}\,\, A^I \Box A_I\,.
\ee
We can study the dynamics of the system in the Hamiltonian formulation,  where the initial data is specified on a constant ``time'' $s=0$ null plane $\Sigma$~(Fig. \ref{fig:null}). The surface element is given by
\be
d\omega_s = \frac{1}{3!}\epsilon_{s\nu \sigma \lambda} dx^\nu dx^\sigma dx^\lambda = d\rho\, dx\, d\bar{x} \,.
\ee
 From the light-cone action, we can define conjugate momenta for the fields $A^I$ 
\be
\pi_I (\mathbf{x}) =\frac{\delta \mathcal{L}}{\delta (\del_s A^I)} = \del_{\rho} A_I\,.
\ee
Note that in the component form, this reads
\be
\pi_A = \del_\rho \bar{A} \,, \quad \pi_{\bar A} = \del_\rho A\,.
\ee
Thus the conjugate momenta are not independent and can be expressed in terms of the $\del_\rho$ derivatives of the fields.
Therefore, the fundamental Poisson brackets are
\bea
&&\{ A_I (\mathbf{x}_1)\, ,A^J(\mathbf{x}_2)\} \Big|_{s_1=s_2}= \frac{1}{\del_\rho}\delta(\rho_1- \rho_2)\, \delta_I{}^J\, \delta^{(2)}(x_1 - x_2) \,.
\eea
Note that the light-cone Minkowski metric is completely off-diagonal, \ie, $ \delta_x{}^{\bar x} = \delta_{\bar x}{}^x = 1$. Hence, the only non-vanishing Poisson brackets is the one between $A(\mathbf{x})$ and $\bar A(\mathbf{x})$.\\

The Poincar\'e transformations on the fields can be obtained from the Lie derivative
\be
\delta_\xi A^\mu = \mathcal{L}_\xi A^\mu = \xi^\nu \del_\nu A^\mu + \del_\nu \xi^\mu A^\nu \,,
\ee
where the parameters have the form $\xi^\mu =\omega^\mu{}_\nu x^\nu +a^\mu$. The precise form of the Poincar\'e transformations in light-cone coordinates is discussed in Appendix~\ref{appendix:A}. We can define canonical generators for Poincar\'e in the phase space of $lc_2$ electromagnetism as
\be
G[\xi] = \int_\Sigma d\omega_s\, \pi_I \delta_\xi A^I = \int_{\Sigma} d \rho dx d\bar{x}\, \del_\rho A_I (\delta_\xi A^I) \,,
\ee
which can be expressed as
\be
G[\xi] =-a_\rho H+ a_s P^s + a_I P^I + \omega_{s\rho} J^{s\rho} + \omega_{sI}J^{sI} + \omega_{x\bar x}J+ \omega_{\rho I} J^{\rho I} \,.
\ee
\par
The explicit form of the light-cone Poincar\'e generators for electromagnetism are listed below.
\begin{itemize}
\item Hamiltonian
\bea
H=- P^\rho &=& \frac{1}{2}\int_{\Sigma}  d \rho dx d\bar{x}\, \del_\rho A_I \del_sA^I =
\frac{1}{2} \int_{\Sigma}  d \rho dx d\bar{x}\, \del_\rho A_I  \frac{\triangle}{\del_\rho}A^I \,,
\eea
\item Momenta
\bea
P^s ~=~ \int_{\Sigma}  d \rho dx d\bar{x}\, \del_\rho A_I \del_\rho A^I\,,&& P^I~=~ \int_{\Sigma}  d \rho dx d\bar{x}\, \del_\rho A_J \del^I A^J \,,
\eea
\item Angular momenta
\bea
J^{s\rho} &=& \int_{\Sigma}  d\rho dx d\bar x  \del_\rho A_I (\rho \del_\rho A^I) \,,\\
J^{sI} &=& \int_{\Sigma}  d\rho dx d\bar x  \del_\rho A_I (x^I\del_\rho A^I) \,, \\
J =J^{x\bar x} &=& \int_{\Sigma}  d\rho dx d\bar x  \del_\rho A_I (\eta^{KL}x_{[K}\del_{L]} - \lambda )A^I \,,
\eea
\item Boosts
\bea
J^{\rho I}& =& \int_{\Sigma}  d\rho dx d\bar x \, \del_\rho A_J [(\rho \del^I - x^I \del^\rho)A^J + \delta_{spin}A^J]\nn \\
&=& \int_{\Sigma}  d\rho dx d\bar x \del_\rho A_J (x^I H + \rho \del^I A^J + \lambda \frac{\del^I}{\del_\rho}A^J)\,,
\eea
\end{itemize}
where $\lambda$ denotes the helicity of the field and $\delta_{spin} A^I$ are the spin correction terms required by the closure of the light-cone Poincar\'e algebra~\cite{Bengtsson:1983pd}. 
\par
The Hamiltonian and boosts involve the time derivative $\del_s$ and thus, behave as dynamical generators, $\mathcal D$ that take us forward in time. The rest of the generators, that do not involve any time derivatives, form the kinematical part $\mathcal K$ of the Poincar\'e algebra that preserve the $s=\text{constant}$ surface. The light-cone Poincar\'e algebra can then be expressed in a compact way as follows.
\bea
&&\mathcal D =\{ H =- P^\rho, J^{\rho I}\} \,, \\
&&\mathcal K = \{ P^s, P^I, J^{s\rho}, J, J^{sI}\} \,, \\ 
&&\{ \mathcal K, \mathcal K \} = \mathcal K\,,\quad \{ \mathcal D, \mathcal K\} = \mathcal D\,, \quad \{ \mathcal D, \mathcal D\} = 0 \,.\label {LC Poincare}
\eea
 Unlike in the instant form case, where the kinematical part $\mathcal K$ is six-dimensional, the light-cone Poincar\'e algebra has seven kinematical generators, which is a key feature of Dirac's front form~\cite{Dirac:1949cp}.
%***************************************************************************************************************************************

\subsection{Residual gauge symmetry}

We now look for the residual gauge symmetries in  $lc_2$ electromagentism
\be
A^\mu  (\mathbf{x})\rightarrow A^\mu  (\mathbf{x}) +  \del^\mu \varepsilon (\mathbf{x}) \,, 
\ee
which preserves the gauge choice $A^s = 0$. The condition $\delta_{\varepsilon} A^s = 0$ constrains the gauge parameter $\varepsilon$ as follows
\be
\del_\rho \varepsilon = 0 \quad \Rightarrow \quad \varepsilon  = \varepsilon (s, x^I)\,.
\ee
These transformation must also satisfy~\eqref{Aminus} such that the solution space remains the same
\be
\quad -\del_s \varepsilon (s, x^I)=  \rho\,  \triangle \varepsilon (s,x^I )\,.
\ee
Since $\varepsilon$ does not depend on $\rho$, this equation can be true only if both the sides vanish identically. Therefore, we get two more constraints on $\varepsilon$
\bea
\dot{\varepsilon} &=& 0\,, \\
\triangle \varepsilon &=&2\del \bar{\del} \varepsilon = 0\,. \label{Laplace}
\eea
The first relation forces the gauge parameter to be time-independent, \ie , $\varepsilon = \varepsilon (x^I)$.
Then, the second condition is simply the Laplace's equation in complex coordinates $(x, \bar x)$, which has the most general solution
\be
\varepsilon(x, \bar x) = P(x) +Q(\bar x) \,.
\ee
Hence, we do not obtain the full residual gauge symmetry parameterized by an arbitrary gauge function $\varepsilon$ of $x^I$. Setting the integration constants to zero in the $lc_2$ formalism, thus, amounts to partially fixing the residual gauge freedom of the theory. The residual gauge parameter in $lc_2$ electromagnetism can only be the sum of a holomorphic and an anti-holomorphic function. Any gauge transformation with an arbitrary parameter $\varepsilon(x, \bar x)$ will take us out of the phase space of $lc_2$ electromagnetism.
\par
The canonical generator of residual gauge transformations are
\bea
G [\varepsilon] = \int_{\Sigma}  d\rho dx d\bar x \, \del_\rho A_I \del^I \varepsilon = \int_{\del \Sigma}  dx d\bar{x} A_I \del^I \varepsilon \,.
\eea
The bracket of $G[\varepsilon]$ with the fields yield the correct transformation law
\be
\{ G[\varepsilon]\, , \,A^I (\mathbf{x})\} = \del^I \varepsilon (\mathbf{x}) \,.
\ee
Since $A^I \rightarrow \mathcal O(\rho^{-1})$, the gauge charge is actually zero. So this subclass of residual gauge transformations would correspond to a trivial or a ``proper'' gauge transformations, that do not affect the physical configuration of the states. Therefore, the residual gauge symmetry in the $lc_2$ formalism is \textit{not} the large gauge transformations of electromagnetism.

%***************************************************************************************************************************************
\section{Light-cone electromagnetism beyond $lc_2$}

%***************************************************************************************************************************************
\subsection{Relaxed boundary conditions}

We now wish to relax the boundary conditions considered in the previous section, so as to recover the full set of residual gauge transformations. We add an $\mathcal O(1)$ term in the expansion of the fields
\bea
\tilde{A}^\mu (\mathbf{x}) &=& V^\mu (s,x^J) + \frac{A^\mu_{(0)} (s, x^J)}{\rho} + \frac{A^\mu_{(1)} (s, x^J)}{\rho^2} + \ldots \, .
\eea
These relaxed boundary conditions are consistent with the action, as the new functions $V^\mu$ do not lead to any divergences in the action. But, one immediately runs into a problem with the boost generator $J^{\rho I}$, as we get some potentially divergent contributions at the leading order
\be
    J^{\rho I} =\int_{\Sigma}  \frac{ d\rho}{\rho} dx d\bar x \,A^J_{(0)}  (x^I \del^K\del_K )V_J \,.
    \ee
    However, if we further restrict the arbitrary functions $V_\mu$ to be of the form of a gauge transformation
    \be
    V_\mu = \del_\mu \Phi (s, x^J)\,,
    \ee
    then the divergences can be eliminated through integration by parts over $x^I$ and using~\eqref{boost-inv}. The total derivative terms with respect to $x^I$ do not contribute as we assume that the fields vanish as $x^I$ approaches infinity. As alluded to earlier, the condition~\eqref{boost-inv} is crucial for ensuring that the light-cone boost generators are well-defined.
    \par
   It is interesting to note here that in the Hamiltonian analysis of asymptotic symmetries at spatial infinity~\cite{Henneaux:2018gfi,Henneaux:2018cst}, one obtains a similar result, \ie, for the canonical generator of Lorentz boosts to be well-defined, the $\mathcal O(1)$ term in the relaxed boundary conditions must be of the form of a gauge transformation.
    \par 
    Henceforth, we shall consider the following boundary conditions 
    \be
    A^I (\mathbf{x}) = \del^I \Phi (s,x^J) + \frac{A^I_{(0)} (s, x^J)}{\rho} + \frac{A^I_{(1)} (s, x^J)}{\rho^2} + \ldots  \,.
\ee
We can also assume a similar fall-off for the $A^\rho$ component
\be
A^\rho (\mathbf{x}) = V^\rho (s,x^J) + \frac{A^\rho_{(0)}(s, x^J)}{\rho} + \frac{A^\rho_{(1)} (s, x^J)}{\rho^2}+ \ldots\,,
\ee
but we will use the constraint equation to determine $A^\rho$ in terms of $A^I$ as was done in the previous section.

%*************************************************************************
\subsection*{Solution space with relaxed boundary conditions} 
The above boundary conditions can be repackaged into a piece that decays as $\rho^{-1}$ or faster, and an $\mathcal O(1)$ term
\be  \label{relaxedBC}
\tilde{A}^I( \mathbf{x})= \del^I \Phi (s, x^J) + A^I (\mathbf{x}) \;, \quad \tilde{A}^\rho( \mathbf{x})= V^\rho (s,x^J) + A^\rho (\mathbf{x}) \,.
\ee
This enables us to freely perform integration by parts on $A^I$ without leading to any surface terms because this piece satisfies the KS boundary conditions.
\par
With these relaxed boundary conditions~\eqref{relaxedBC}, the constraint equation~\eqref{constraint} reads 
\bea \label{constraint-relaxedBC}
 \tilde{A}^\rho &=&  -\frac{\del_I {A}^I}{\del_\rho} + (\triangle \Phi -\alpha) \rho +\beta \,.
\eea
One can then identify $V^\rho$ with $\beta$. For finiteness, we demand $\tilde{A}^\rho  \sim \mathcal O(1)$. Thus, at the leading order, the constraint reduces to 
\be
\alpha = \triangle \Phi \,.
\ee
At the subleading orders, we recover the old constraint relations as in the previous section
\be 
\del_I A^I_{(0)} (s, x^J)= 0 \,,
\ee 
\be \label{constraint2}
 A^\rho_{(n-1)} (s, x^J)\, = \frac{1}{(n+1)}\del_I A^I_{(n)} (s, x^J)\,,  \quad n = 1,2,... \, .
\ee
\par
The dynamical equation now becomes
\bea \label{dynamical}
&& \Box \tilde{A}^I - \del^I \alpha = 0 
 \quad \Leftrightarrow \quad  \Box A^I = 0 \,,
 \eea
 indicating that the piece in $\tilde{A}^I$ that satisfy the KS boundary conditions, namely $A^I$, are the propagating degrees of freedom as before.
The trivial equation~\eqref{trivial} is no longer an identity but yields a relation between the constants $\alpha$ and $\beta$
\bea \label{trivial}
\del_s \alpha = \triangle \beta\,,
 \eea
implying that only one of them is arbitrary. Henceforth, we shall assume that $\alpha$ is arbitrary and $\beta$ is determined through the above relation.

%***************************************************************************************************************************************
\subsection{Modified action for electromagnetism}
For a general set of boundary conditions, where the solution space includes the constants $\alpha$ and $\beta$, the light-cone action for electromagnetism is modified as follows
\bea
\mathcal{S} [\tilde{A}^I , \alpha] = \int d^4 \mathbf{x} \left( \frac{1}{2}\tilde{A}^I \Box \tilde{A}_I + \alpha\, \del_I \tilde{A}^I - \frac{1}{2}\alpha^2\right) + B_{\infty} \,,
\eea
where the boundary term $B_\infty$ to evaluated in the limit $\rho \rightarrow \pm \infty$ is given by
\bea
B_\infty = \int ds dx d\bar{x}\, (\tilde{A}^I \del_s \tilde{A}_I +\tilde{A}^\rho \del_\rho \tilde{A}^\rho) \,.
\eea
One can also derive this action from a covariant Lagrangian with a gauge-fixing term as shown in appendix~\ref{appendix:B}.
\par
With the relaxed boundary conditions $\tilde{A}^I = \del^I \Phi + A^I$ and the corresponding value of the constraint $\alpha= \triangle \Phi$, the modified action reads
\bea
\mathcal S[A^I, \Phi] &=& \frac{1}{2}  \int ds d\rho dx d\bar{x}\, A^I \Box A_I \,+\, B_\infty \,,
\eea
\vskip 0.2cm
\ndt
with the boundary term
\bea
B_{\infty} &=& - \int ds dx d\bar{x}\,\dot{\Phi} \triangle \Phi  \,.
\eea
Thus, we find that the bulk action is unchanged but the new field $\Phi$ plays the role of a surface degree of freedom which induces a boundary term in the action. This implies that the phase space is now enlarged to $\{A^I, \Phi\}$ with a new  ``equal-time'' Poisson bracket on the two-dimensional surface
\be
 \{ \Phi(x)\, ,\ \triangle \Phi(x')\}\Big|_{s=s'}~=~ \delta^{(2)} (x-x')\,.
\ee
As we have seen for the fields $A^I$, the conjugate momentum for $\Phi$ is also not independent, which is an artefact of the light-front Hamiltonian formulation.

%***************************************************************************************************************************************
\subsection{Residual gauge symmetry}
We can now return to the question of residual gauge transformations that leaves invariant the modified light-cone action
\bea
\mathcal{S} [\tilde{A}^I , \alpha] = \int d^4 \mathbf{x} \left( \frac{1}{2}\tilde{A}^I \Box \tilde{A}_I + \alpha\, \del_I \tilde{A}^I - \frac{1}{2}\alpha^2\right) + B_{\infty} \,.
\eea
As before, the gauge parameter must be independent of $\rho$ in order to preserve the light-cone gauge condition $A^s =0$
\be
 \epsilon (\mathbf{x}) = \varepsilon(s, x^I)\,.
\ee
We find that the set of gauge transformation 
\be
\delta_\varepsilon \tilde{A}^I = \del^I \varepsilon\, , \quad \delta_\varepsilon\alpha =  \triangle \varepsilon \,,
\ee
is now a symmetry of the modified light-cone action without the additional Laplace's equation~\eqref{Laplace}.  
\par
In terms of the dynamical fields $A^I$ and the boundary mode $\Phi$
\bea
\mathcal S[A^I, \Phi] &=& \frac{1}{2} \int ds d\rho dx d\bar{x}\,  A_I \Box A^I  -  \int ds dx d\bar{x} \, \dot{\Phi} \triangle \Phi \,,
 \eea
we find that the residual gauge transformations read 
\be
\delta_\varepsilon A^{I} =0\, , \quad \delta_\varepsilon \Phi  (s, x^I)= \varepsilon (s, x^I) \,.
\ee
Therefore, these residual gauge transformations act as a shift symmetry on $\Phi$, while leaving $A^I$ unchanged. This explains why the large gauge transformations are absent in the $lc_2$ formalism. In order to ``see'' the large gauge transformations in light-cone electromagnetism, we must keep the boundary field $\Phi$. This implies that going from $lc_2$ electromagnetism to $lc_4$, where one keeps the integration constants $\alpha$ and $\beta$, involves a large gauge transformation. The complete set of all residual gauge transformations in light-cone electromagnetism are
\begin{itemize}
\item Proper: $\delta_{\varepsilon} A^I = \del^I \varepsilon (s, x^J) \;, \delta_{\varepsilon} \Phi = 0$ with $\triangle \varepsilon =0\,,$ 
\item Improper: $\delta_{\varepsilon} A^I =0\;, \delta_{\varepsilon} \Phi = \varepsilon (s, x^J)$ with $\triangle \varepsilon \neq 0\,.$
\end{itemize}
\par
It is worthwhile to mention that in case of light-cone gravity, in order to obtain the BMS supertranslations, which act as large or improper diffeomorphisms, one does not need to introduce any extra degrees of freedom~\cite{Ananth:2020ojp}. The supertranslations can be realised without extending the phase space beyond that of $lc_2$ gravity. Thus, the analysis of residual gauge symmetry in the light-cone formulation shares striking similarities with the asymptotic structure at spatial infinity as one finds the same puzzling feature in these analyses~\cite{Henneaux:2018gfi, Henneaux:2018cst}.
%***************************************************************************************************************************************
\subsection{Canonical generators and symmetry algebra}
The canonical generator of residual gauge transformations now becomes
\bea
G [\varepsilon] = \int_\Sigma d\rho dx d\bar x \, \del_\rho A_I \del^I \varepsilon +\mathcal Q[\varepsilon] \,,
\eea
where the non-vanishing charge for large gauge transformations only involves $\Phi$
\be
\mathcal Q[\varepsilon] =  \int_{\del \Sigma} dx d\bar{x} \,\triangle \Phi \varepsilon \,.
\ee
We can check that the bracket of $G[\varepsilon]$ with the fields yield the correct transformation law
\bea
\{ A^I (\mathbf{x})\,, G[\varepsilon]\, \} &=& \del^I \varepsilon (\mathbf{x}) \;, \\
 \{ \Phi (\mathbf{x})\,,G[\varepsilon]\, \} &=&\varepsilon (\mathbf{x})  \,.
\eea
\par
We can compute the surface contributions coming from $\Phi$ to the canonical generators for Poincar\'e transformations. The Poincar\'e transformation of the scalar $\Phi$ reads
\be
\delta_\xi \Phi = \xi^\nu \del_\nu \Phi \,.
\ee
Since $\Phi$ does not depend on $\rho$, the generators involving $\del_\rho$ have no surface corrections. 
The surface corrections to the other Poincar\'e generators listed in Section~(\ref{section:2.3}) are as follows
\be
\mathcal Q[\xi] = a_\rho H + a_I P^I + \omega_{\rho I} J^{\rho I} + \omega_{x\bar x} J \,,
\ee
where
\bea
H= -P^\rho = \int_{\del \Sigma} dx d\bar x\, \triangle \Phi \dot{\Phi} \,,&& P_I = \int_{\del \Sigma} dx d\bar x \,\triangle \Phi \del_I \Phi \,,  \\
J^{\rho I} = \int_{\del \Sigma} dx d\bar x \,\triangle \Phi (x^I \dot{\Phi}) \,, && J = \int_{\del \Sigma} dx d\bar x\, \triangle \Phi ( x^{[I} \del^{J]} \Phi) \,.
\eea
\par
With these generators, we find that the asymptotic algebra of light-cone electromagnetism is given by
\bea
\{G[\xi_1], G[\xi_2] \} &=& G[\hat{\xi}] \,, \\
\{G[\varepsilon_1], G[\varepsilon_2] \} &=&0 \,, \\
\{G[\varepsilon_1],  G[\xi]  \} &=& \mathcal Q[\hat{\varepsilon}] \,.
\eea
The first bracket stands for the light-cone Poincar\'e algebra given in~\eqref{LC Poincare}, where the generators have been augmented with the surface corrections. The second bracket shows that the gauge algebra is  Abelian as expected\footnote{In presence of sources or magentic monopoles, the gauge algebra might contain a central charge~\cite{Freidel:2018fsk}. But we assume that all such sources are absent in our analysis.}. The third bracket tells us that under a Poincar\'e transformation the gauge parameter becomes
\be
\hat{\varepsilon} = \xi^\mu \del_\mu \varepsilon\,.
\ee
Thus, we find that with the relaxed boundary conditions we can recover the large gauge transformations in light-cone electromagnetism, that are parameterized by an arbitrary function $\varepsilon$.

%***************************************************************************************************************************************

\section{Newman-Penrose formalism}
In this section, we employ Newman-Penrose formalism~\cite{Newman:1961qr} to identify the incoming and outgoing radiation fields in light-cone electromagnetism. The Minkowski metric in light-cone coordinates is naturally endowed with a set of four null tetrads that form a basis for the Newman-Penrose formalism.
\par
 The light-cone Minkowski metric may be recast into the tetrad form
\be
\eta_{\mu\nu} = e^{(a)}_\mu e^{(b)}_\nu \eta_{(a)(b)} \,,
\ee
where the local Lorentz indices $(a), (b),\ldots$ run over $(s,\rho, x, \bar x)$. The metric of the internal space is also expressed in light-cone coordinates. The details of our notations are presented in Appendix~\ref{appendix:A}. \\

We define the Newman-Penrose (NP) null tetrads as follows 
\bea
\mathit{l} = \mathit{e}_{(s)} = - \mathit{e}^{(\rho)} \;, \quad \mathit{n} = \mathit{e}_{(\rho)} = - \mathit{e}^{(s)} \;, \quad \mathit{m} = \mathit{e}_{(x)} =  \mathit{e}^{(\bar x)} \;, \quad \mathit{\bar{m}} = \mathit{e}_{(\bar x)} =  \mathit{e}^{(x)}\,.
\eea
In component form, the tetrads read
\bea
l^\mu &=& (1,0,0,0) \\
 n^\mu &=& (0,1,0,0)  \\
 m^\mu &=& (0,0,0,1) \\
 \bar{m}^{\mu}&=& (0,0,1,0)\
\eea
One can check that the set of tetrads are null
\bea
\mathit{l}.\mathit{l} = \mathit{n}.\mathit{n} =\mathit{m}.\mathit{m} =\mathit{\bar{m}}.\mathit{\bar{m}} =0 \,,
\eea
and that they satisfy
\be
\mathit{l}.\mathit{n}= \mathit{n}.\mathit{l} = -1\, , \quad \mathit{m}.\mathit{\bar{m}}=\mathit{\bar{m}}.\mathit{m} = 1 \,.
\ee
Note that our convention differ by a sign from the ones originally introduced by Newman and Penrose. The derivative operator may be expressed in the basis of the NP tetrads as
\be
\del_\mu = - l_\mu \boldsymbol{\Delta}-n_\mu \textbf{D} + m_\mu\boldsymbol{\bar{\delta}} + \bar{m}_\mu \boldsymbol{\delta}\,,
\ee
where the symbols stand for the directional derivatives~\footnote{The directional derivative denoted by the boldface symbol $\boldsymbol{\Delta}$ is not to be confused with the two-dimensional Laplacian operator $\triangle =\del_I \del^I$.}
\bea
\boldsymbol{\Delta}=\del_\rho \;, \quad  \boldsymbol{D} = \del_s \, , \quad \boldsymbol{\bar{\delta}} =\bar{\del} \;,\quad \boldsymbol{\delta} ={\del} \,.
\eea
\par
In light-cone electromagnetism, the gauge condition on the vector field $A^\mu = (A^s, A^\rho, A, \bar A)$ can be implemented as
\be
\mathit{n}. \mathit{A} = -A^s = A_\rho= 0 \,.
\ee
The field strength $F_{\mu\nu}$ is given in terms of the three complex Newman-Penrose scalars defined as follows
\bea
&&\phi_0 = F_{\mu \nu} \,l^\mu n^\nu = \del_s  A + {\del}A^\rho \,,\\
&& \phi_1 = \frac{1}{2} F_{\mu \nu} (l^\nu n^\nu + \bar{m}^\mu m^\nu) = \frac{1}{2}(\del_\rho A^\rho + \bar{\del}{A} - {\del}\bar{A}) \,, \\
&& \phi_2 = F_{\mu \nu}\bar{m}^\mu n^\nu=  -\del_\rho \bar{A} \,.
\eea
The free Maxwell equations in Minkowski in terms of the NP scalars read
\bea
\boldsymbol{D}\phi_1 - \boldsymbol{\bar{\delta}} \phi_0 &=& 0 \,,\\
\boldsymbol{D}\phi_2 - \boldsymbol{\bar{\delta}} \phi_1 &=& 0 \,, \\
\boldsymbol{\Delta}\phi_0 - \boldsymbol{\delta} \phi_1 &=& 0 \,, \\
\boldsymbol{\Delta}\phi_1 - \boldsymbol{\delta} \phi_2 &=& 0\,. 
\eea
One can check that the fourth equation corresponds to the familiar light-cone constraint equation ~\eqref{constraint-relaxedBC}, the second and third equations to the dynamical equations~\eqref{dynamical} for $\bar{A}$ and $ A$ respectively, and the first equation replaces the trivial equation~\eqref{trivial}. 
\par 
With the relaxed boundary conditions introduced in the previous section, 
\bea
\tilde{A}  = \del \Phi + A\;, \quad \tilde{\bar{A}}  = \bar{\del} \Phi + \bar{A} \quad \text{with} \quad  (A, \bar{A} )\sim \mathcal O\left( \frac{1}{\rho}\right) \,,
\eea
we can compute the NP scalars that satisfy the above equations of motion
\bea
\phi_0 &=& {\del} \dot{\Phi}+ \triangle A_{(0)}+\mathcal O\left( \frac{1}{\rho}\right)\,, \label{phi0}\\
\phi_1 &=&  \frac{{\del} \bar{A}_{(0)}}{\rho} + \mathcal O \left( \frac{1}{\rho^2} \right)\,, \label{phi1}\\
\phi_2 &=& \frac{\bar{A}_{(0)}}{\rho^2} + \mathcal O \left( \frac{1}{\rho^3} \right)\,.\label{phi2}
\eea
Here, in computing $\phi_0$ we have used the condition $\del \bar{A}_{(0)}+ \bar{\del}A_{(0)} =0$ imposed by boost invariance~\eqref{boost-inv}.
\par
In the Newman-Penrose formalism, the Maxwell scalars $\phi_0$ and $\phi_2$ represent the incoming and outgoing radiation fields respectively. Indeed, we see that in light-cone electromagnetism, the scalar $\phi_0$ which has helicity 
plus one involves $A$, while $\phi_2$ with helicity minus one only contains $\bar{A}$. These fields truly behave like the two massless helicity states of the photon that carry electromagnetic radiation.
\par
We can also derive the Poincar\'e charges in the Newman-Penrose formalism from the energy-momentum tensor of the Maxwell field~\cite{Newman:1961qr, Chandrasekhar:1985kt}
\be \label{energy-mom}
T_{\mu \nu} = \eta^{\lambda \sigma} F_{\mu \lambda} F_{\nu \sigma} - \frac{1}{4} \eta_{\mu \nu} F_{\lambda \sigma}F^{\lambda \sigma} \,,
\ee
which in terms of the Maxwell field reads 
\bea
-\frac{1}{2} T_{11} = \phi_0 \phi_0{}^*\,, && -\frac{1}{2}T_{13} = \phi_0 \phi_1{}^* \,,\\
-\frac{1}{4} (T_{12}+ T_{34}) = \phi_1 \phi_1{}^*\,,&& -\frac{1}{2}T_{23} = \phi_1 \phi_2{}^*\,,\\
-\frac{1}{2}T_{22} = \phi_2 \phi_2{}^* \,, && -\frac{1}{2}T_{33} = \phi_0 \phi_2{}^*\,.
\eea
The light-cone Poincar\'e generators are then given by 
\bea
P^\mu &=& \int_\Sigma d\rho dx d\bar x \, n_\mu T^{\mu \nu} \,,\\
J^{\mu \nu} &=& \int_\Sigma d\rho dx \bar x  \, n_\lambda (x^\mu T^{\lambda \nu} - x^\mu T^{\lambda \mu}) \,.
\eea
However, we shall not derive the Poincar\'e generators again as these have been discussed in detail in the previous sections.

%***************************************************************************************************************************************
\subsection{Quadratic form Hamiltonians}
In the $lc_2$ formalism, the Hamiltonian can often be shown to possess a \textit{positive definite} quadratic form structure. This recurring structure appears in spin one theories, namely electromagnetism, Yang-Mills and $\mathcal N=4$ super Yang-Mills theory~\cite{Ananth:2015tsa}. For spin two, this structure is found in Einstein's gravity and maximal supergravity in various dimensions~\cite{Ananth:2006fh, Ananth:2017xpj}. In a recent work, the quadratic form structure has been shown to exist in higher spin theories as well~\cite{Ananth:2020mws}.  It has been argued that the quadratic form structure of the Hamiltonian or energy density is related of the positivity of energy in light-cone theories. 
\par
In general, the quadratic form Hamiltonians in $lc_2$ theories in four dimensions have the form
\be
H = \int_\Sigma d\rho dx d\bar x \, \mathcal H [\Psi, \overline{\Psi}] \,,
\ee
with the Hamiltonian density
\be \mathcal H [\Psi, \overline{\Psi}] ~=~  \left( \mathcal D \overline{\Psi} \right) \overline{\mathcal D} \Psi \,,
\ee
where the fields $\Psi$ and $\overline{\Psi}$, respectively stand for the positive and negative helicity states of the massless particles in the theory. The operator $\mathcal D$, referred to as the ``covariant derivative'' in the $lc_2$ literature, has a highly non-linear structure which is theory-dependent. In Table~\ref{tab:QF}, we summarize the quadratic form structure of light-cone Hamiltonians for some non-supersymmetric theories.
\begin{center}
\begin{table}[h!] 
\caption{\textbf{Quadratic form Hamiltonians of light-cone theories with massless fields}\\
The boldface subscripts in square brackets denote the helicity of the field; the latin indices $a, b,c, ...$  belong to the $SU(N)$ gauge group; the symbols $g,\kappa, \alpha$ are the coupling constants.}
\begin{tabular}{l l c}
\label{tab:QF}
\textbf{Theory}&  \textbf{ ``Covariant derivative''} $\overline{\mathcal{D}} \Psi $& \textbf{Hamiltonian $\mathcal H$}\\ 
 \hline \\
\textbf{Maxwell} & &\\
$A_{[\mathbf{1}]},\bar{A}_{[\mathbf{-1}]} $& $\overline{\mathcal D}  A = \bar{\del} A$ &$ \mathcal D \bar{A} \,\overline{\mathcal D}  A  $ \\\\
\hline\\
\textbf{Yang-Mills} & &\\
 $A^a_{[\mathbf{1}]},\bar{A}^a_{[\mathbf{-1}]} $ & $\overline{\mathcal D}  A^a = \bar{\del} A^a + g\, f^{abc}\frac{1}{\del_\rho}(\bar{A}^b\del_\rho A^c)$ & $ \mathcal D \bar{A}^a \overline{\mathcal D}  A^a  $\\\\
 \hline\\
 \textbf{Gravity} & &\\
 $h_{[\mathbf{2}]},\bar{h}_{[\mathbf{-2}]} $& $\overline{\mathcal D}  h = \bar{\del} h + 2\kappa \frac{1}{\del_\rho^2} \left( \frac{\del}{\del_\rho}\bar{h} \del_\rho^3 h - \bar{h} \del_\rho^2 \del h \right) + \mathcal{O}(\kappa^2)$   & $ \mathcal D \bar{h} \overline{\mathcal D}  h $    \\\\
 \hline\\
  \textbf{Higher spins} & &\\
 $\varphi_{[\boldsymbol{\lambda}]},\bar{\varphi}_{[\boldsymbol{-\lambda}]} $& $\overline{\mathcal D}  \varphi =\bar{\del} \varphi - 2\alpha \sum_{n=0}^{\lambda-1}(-1)^n {{\lambda-1 }\choose {n}} \frac{\bar{\del}^n}{\del_\rho^{n+1}} \left[ \frac{\del^{\lambda -n-1}}{\del_\rho^{\lambda-n-1}} \bar{\varphi} \del_\rho^\lambda \varphi \right]$ & $ \mathcal D \bar{\varphi} \overline{\mathcal D}  \varphi $ \\\\
\hline
\end{tabular}
\end{table}
\end{center}
For $lc_2$ electromagnetism, the quadratic form Hamiltonian reads
\be
H =  \int_\Sigma d\rho dx d\bar x\, \del A \bar{\del}A \,.
\ee
In the Newman-Penrose scheme, the Hamiltonian in light-cone coordinates, $H = -P^\rho$, is given by 
\be
H = \frac{1}{2} \int_\Sigma d\rho dx d\bar x\, T^{\mu \nu}  n_\mu l_\nu ~=~ \frac{1}{4}  \int_\Sigma d\rho dx d\bar x\,T^{\mu \nu} (n_\mu l_\nu + \bar{m}_\mu {m}_\nu) \;,
\ee
where the second equality follows from the tracelessness of the energy-momentum tensor. With the energy momentum tensor for electromagnetism~\eqref{energy-mom}, one can show that the Hamiltonian becomes
\be \label{NP-QF}
H= \int_\Sigma d\rho dx d\bar x \, \phi_1 \phi_1{}^* \,.
\ee
Hence, the covariant derivative appearing in the quadratic form Hamiltonian is actually one of the Newman-Penrose scalars. This seems to imply that in a given $lc_2$ theory, the ``covariant derivative'' $\overline{\mathcal D}\Psi$  that appears in the quadratic form Hamiltonian is the Newman-Penrose (NP) scalar defined by projecting the field strength (or curvature tensor for spin two) along the two-form
\be
\Sigma_{\mu\nu}= \frac{1}{2}(n_\mu \wedge l_\mu - m_\mu \wedge \bar{m}_\mu) \,.
\ee
\par
 The case for electromagnetism may seem trivial as the scalar is simply given by~\eqref{phi1}
\be
\phi_1 = \bar{\del} A\,.
\ee
But, one can also check this ``conjecture'' for the non-Abelian case, namely $lc_2$ Yang-Mills theory in four dimensions. The corresponding scalar is defined as
\be
\phi^a_1= \frac{1}{2} F^a_{\mu \nu} (l^\mu n^\nu + \bar{m}^\mu m^\nu ) \,,
\ee
where $F^a_{\mu \nu}$ is the non-Abelian field strength given by
\be
F^a_{\mu \nu} = \del_\mu A_\nu - \del_\nu A_\mu + g f^{abc} A^b A^c \,.
\ee
Working in the light-cone gauge $A^{as}= - A^a_\rho = 0$, we can solve for $A^{a\rho}$ in terms of $A^{aI}$ from the constraint equation. We find that the scalar $\phi^a_1$ then reads
\be
\phi^a_1 = \bar{\del} A^a + g\,f^{abc}\frac{1}{\del_\rho} (\bar{A}^b\del_\rho A^c) \,.
\ee
When substituted the expression for the Hamiltonian in terms of the NP scalar
\be
H = \int_\Sigma d\rho dxd\bar x\, \phi^a_1 \phi^a_1{}^* \,,
\ee
this indeed reproduces the quadratic form structure for Yang-Mills as depicted in Table~(\ref{tab:QF}).
\par
For gravity, this suggests that the covariant derivative $\overline{\mathcal D} h$ in the light-cone Hamiltonian is related to the scalar $\phi_{11}$ in the Newman-Penrose formalism for spin two. In a recent work~\cite{Ananth:2020ojp}, we had commented on possible interpretations of the covariant derivative for light-cone gravity in the context of gravitational radiation and BMS symmetry. A similar Newman-Penrose analysis for light-cone gravity can shed some light on the question whether this feature, namely the covariant derivative $\overline{\mathcal D} \Psi$ is a NP scalar, is shared by all massless $lc_2$ theories or a mere coincidence for spin one theories.

%***************************************************************************************************************************************

\section{Concluding remarks}
Based on the $(3+1)$ Hamiltonian formulation, the asymptotic analysis at spatial infinity \`a la Henneaux-Troessaert~\cite{Henneaux:2018gfi, Henneaux:2018cst} exhibits some interesting features that also appear in the light-cone analysis, such as the structure of the phase space, subtleties involving boost generators and the symmetry algebra of canonical generators. The light-cone gauge-fixed action is essentially a first order system in time derivatives, which could explain why our analysis of residual gauge symmetries is so similar to the asymptotic symmetry analysis at spatial infinity.  One pressing issue that remains to be understood is why electromagnetism requires a boundary field for the large gauge transformations, while no such extra field is needed for gravity. However, this puzzle may only be investigated with a more rigorous Hamiltonian analysis for light-cone theories where the role of these boundary or zero mode fields become transparent. 
\par
In the light-cone formulation, we usually start from a covariant Lagrangian, impose the light-cone gauge conditions and systematically eliminate the unphysical degrees of freedom by solving the constraints. The light-cone Lagrangian action so obtained can then be used to define conjugate momenta and derive the Hamiltonian using the Legendre transform. But the relation ~\eqref{NP-QF}, if satisfied in general for $lc_2$ theories, offers an alternative route to deriving interacting Hamlitonian action from the equations of motion without the knowledge of the light-cone Lagrangian. One can then apply it to the powerful framework of~\cite{Bengtsson:1983pd} to find all canonical generators of Poincar\'e for a given theory from the closure of the algebra alone.
\par
The fact that the $lc_2$ theories are described solely in a basis of the two helicity eigenstates, makes these theories well-suited for the spinor helicity formalism $-$ the building blocks for modern amplitude techniques~\cite{Ananth:2012un}. The $lc_2$ formalism is also closely related to the self dual and anti self-dual formulation of Yang-Mills theories~\cite{Chalmers:1996rq}.  A Newman-Penrose analysis of $lc_2$ theories might help us appreciate these intriguing connections to a greater extent. The questions pertaining to gravitational radiation in the light-cone formulation might be easier to address by reformulating $lc_2$ gravity in the Newman-Penrose formalism.  Another important direction would be to establish a light-cone double copy construction following the work of~\cite{Campiglia:2021srh}, where we can map the symmetries from gauge theories to gravity. 

\section*{Acknowledgements}
We thank Sudarshan Ananth, Glenn Barnich and Marc Geiller for many fruitful discussions. The author is grateful to the LABEX Lyon Institute of Origins (ANR-10-LABX-0066) Lyon for its financial support within the Plan France 2030 of the French government operated by the National Research Agency (ANR). 
%***************************************************************************************************************************************
\appendix

%***************************************************************************************************************************************
\section{Light-cone coordinates and Poincar\'e transformations}
\label{appendix:A}
\subsection*{Notations and conventions}

The light-cone coordinates are defined as
\be
x^+ = \frac{x^0 +x^3}{\sqrt 2}\,, \quad x^- = \frac{x^0-x^3}{\sqrt 2}\, ,\quad  x= \frac{x^1 + ix^2}{\sqrt 2}\, , \quad \bar{x} = \frac{x^1-ix^2}{\sqrt 2} \,.
\ee
We relabel the light-cone coordinates for our convenience as $x^\mu = (s,\rho, x^I)$
\be
s =x^+\,, \quad  \rho = x^- \, ,\quad x^I \equiv (x, \bar x)\,.
\ee
The Minkowski metric in the light-cone coordinates is
\be
(\eta_{\mu\nu})^{LC} =
\left( \begin{array}{cccc}
0&-1&0&0 \\
-1&0&0&0 \\
0& 0& 0&1 \\
0 & 0 & 1& 0
\end{array} \right)\,.
\ee
The derivative operator $\del_\mu = (\del_s, \del_\rho, \del_I)$ reads
\bea
&&\del_s = \frac{\del_0 + \del_3}{\sqrt 2} \,, \quad \del_\rho = \frac{\del_0 - \del_3}{\sqrt 2} \,, \quad \bar{
\del} = \ \frac{\del_1 - i\del_2}{\sqrt 2} \, ,\quad \del =  \frac{\del_1 + i \del_2}{\sqrt 2} \,,\\
&&\text{such that} \quad \del_I x^I = 2\, ,\quad \text{and} \quad \del_I= (\bar{\del}, \del) \,.
\eea
The d'Alembertian becomes
\be
\Box = \del_\mu \del^\mu = -2 \del_s \del_\rho + \del_I \del^I= -2 \del_s \del_\rho +2 \del \bar{\del}\,.
\ee
Volume element in light-cone coordinates is defined as
\be
d\omega = \frac{1}{4!}\epsilon_{\mu \nu \sigma \lambda} dx^\mu dx^\nu dx^\sigma dx^\lambda = ds\, d\rho\, dx\, d\bar{x} \, , \quad \text{with} \quad \epsilon_{s\rho x \bar{x}} = 1  \,.
\ee
A covariant and contravariant vector in these coordinates have the form
\be
A^\mu = (A^s, A^\rho, A, \bar A) \, \quad \text{and} \quad B_{\mu} = (B_s, B_\rho, \bar{B}, B) \,,
\ee
such that their dot product is given by
\be
A.B = A^\mu B_\mu =  A^s B_s +A^\rho B_\rho + A^I B_I =  - A^s B^\rho - A^\rho B^s + A \bar{B} + \bar{A} B \,.
\ee

%******************************************************************************

\subsection*{Light-cone Poincar\'e transformations}
The Poincar\'e transfromations $x^\mu \rightarrow  x^\mu + \xi^\mu (\mathbf{x})$ with
\be
 \xi^\mu (\mathbf{x}) = \omega^\mu{}_\nu x^\nu + a^\mu\, ,
\ee
in light-cone coordinates $(s, \rho, x^I)$ take the form
\bea
\xi^s& =& \omega_{s\rho} s + \omega_{\rho I}x^I + a^s \,,\\
\xi^\rho & =& - \omega_{s\rho} \rho + \omega_{sI} x^I + a^\rho \,,\\
\xi^I &=& -\omega_{sI} s - \omega_{\rho I} \rho + \omega^I{}_{J} x^J + a^I \,.
\eea
Thus, the light-cone Lorentz transformations are parameterized by $(\omega_{s\rho}, \omega^I{}_J)$, which are real and $(\omega_{sI},\omega_{\rho I})$ , which are complex. The space-time translations are spanned by the real 4-vector $(a^s, a^\rho, a^I)$. 
\par 
Note that since the light-cone Minkowski metric is completely off-diagonal in the two pairs $(s, \rho)$ and $(x, \bar x)$, the Lorentz transformations do not mix $s$ with $\rho$ and $x$ with $\bar x$.
\par
The transformation laws for the fields $A^\mu$ and $\Phi$ can then be obtained in the usual way
\bea
\delta A^\mu &=& \xi^\nu \del_\nu \xi^\mu + \del_\nu \xi^\mu A^\mu\, , \\
\delta \Phi& =& \xi^\nu \del_\nu \Phi \,.
\eea
The precise form of the transformation laws for the fields $A^I$ and $\Phi$ can be found in~\cite{Bengtsson:1983pd}.

%******************************************************************************
\subsection*{Light-cone Minkowski in tetrad form}
 The light-cone Minkowski metric may be recast into
\be
\eta_{\mu\nu} = e^{(a)}_\mu e^{(b)}_\nu \eta_{(a)(b)} \; ; \quad (a), (b),\ldots = (s,\rho, x, \bar x) \,,
\ee
where $(a), (b)$ represent the local Lorentz indices.  We choose the metric on the internal space to be expressed in light-cone coordinates as well
\be \label{local-Lorentz}
 \eta_{(a)(b)} = \left( \begin{array}{cccc}
 0&-1&0&0 \\
-1&0&0&0 \\
0& 0& 0&1 \\
0 & 0 & 1& 0
 \end{array}
 \right)\,,
\ee
such that the tetrads reads
\bea
e^{(s)}_\mu = (0, -1, 0, 0)\,, \quad
e^{(\rho)}_\mu = (-1, 0, 0, 0) \,,\quad 
e^{(x)}_\mu = (0, 0, 1, 0) \,,\quad 
e^{(\bar{x})}_\mu= (0,0,0,1) \,.
\eea
The dual tetrads can be obtained from the metric~\eqref{local-Lorentz}
\be
e _{\mu (a)} = \eta_{(a)(b)} e^{(b)}_{\mu}\ .
\ee
\par
We define the Newman-Penrose (NP) null tetrads as follows
\bea
\mathit{l} = \mathit{e}_{(s)} = - \mathit{e}^{(\rho)} \;, \quad \mathit{n} = \mathit{e}_{(\rho)} = - \mathit{e}^{(s)} \;, \quad \mathit{m} = \mathit{e}_{(x)} =  \mathit{e}^{(\bar x)} \;, \quad \mathit{\bar{m}} = \mathit{e}_{(\bar x)} =  \mathit{e}^{(x)}\,.
\eea
In component form, the tetrads read
\bea
l^\mu = (1,0,0,0)\, , \quad n^\mu = (0,1,0,0) \, ,\quad m^\mu = (0,0,0,1) \, , \quad \bar{m}^{\mu}= (0,0,1,0)\,.
\eea
A 4-vector in the new basis has the following form
\be
P^\mu = \mathit{l}^\mu  P^+ + \mathit{n}^\mu P^- + \mathit{m}^\mu \bar{P} + \mathit{\bar{m}}^\mu P\,.
\ee
\par
Alternatively, one can start from the Minkowski metric in Cartesian coordinates
\be
\eta_{\mu \nu} = diag(-1, 1, 1,1) \; ; \quad \mu, \nu, \ldots = 0,1,2,3,
\ee
and use the tetrads expressed in these coordinates
\be
\mathit{e}_{(0)} = (1,0,0,0) \;,  \quad \mathit{e}_{(1)} = (0,1,0,0) \;,  \quad  \mathit{e}_{(2)} = (0,0,1,0) \;, \quad  \mathit{e}_{(3)} = (0,0,0,1) \,,
\ee
to define the a set of null tetrads for light-cone Newman-Penrose formalism as done in~\cite{Leibbrandt:1984be}
\bea
&&\mathit{l} = \frac{e_{(0)}+ e_{(3)}}{\sqrt 2} \; , \quad  \mathit{n} = \frac{e_{(0)}- e_{(3)}}{\sqrt 2}\; , \\
&&\mathit{m} = \frac{e_{(1)}+i e_{(2)}}{\sqrt 2} \; , \quad  \mathit{\bar{m}} = \frac{e_{(1)}-i e_{(2)}}{\sqrt 2} \,.
\eea

%
%***************************************************************************************************************************************

\section{Modified action from a covariant Lagrangian}
\label{appendix:B}
One can derive the modified light-cone action given in section 3 from a covariant Lagrangian with a gauge-fixing term. Let us consider the Maxwell Lagrangian with a gauge-fixing term
\bea
\mathcal L = -\frac{1}{4} F^{\mu \nu}F_{\mu \nu} - \frac{1}{2\xi} (\del_\mu A^\mu)^2 \,,
\eea
where $\xi$ is a gauge parameter.  In this appendix, we suppress the \textit{tilde} on the fields $A^\mu$ for simplicity.
 Another way to implement this kind of gauge-fixing is through the introduction of an auxiliary field $B$
\bea
\mathcal L = -\frac{1}{4}F^{\mu \nu} F_{\mu \nu} + \frac{1}{2} \xi B^2+ B (\del_\mu A^\mu) \,,
\eea
such that the equation of motion for $B$ enforces the constraint
\be
\xi B + \del_\mu A^\mu = 0 \quad \Rightarrow \quad B = -\frac{1}{\xi} (\del_\mu A^\mu)\,.
\ee
\par
Similarly, the modified action for light-cone electromagnetism with the relaxed boundary conditions 
\be
\mathcal S [A^I, \alpha] = \int d^4x \left\{ \frac{1}{2} A^I \Box A_I + \alpha (\del_IA^I) - \frac{1}{2}\alpha^2\right\} \,,
\ee
may indeed be derived from a covariant gauge-fixed Lagrangian
\be
\mathcal L = -\frac{1}{4} F^{\mu \nu}F_{\mu \nu}+ \alpha (\del_\mu A^\mu)+ \frac{1}{2}\xi \alpha^2 \,,
\ee
where $\alpha$ plays the role of the auxiliary field, whose equation of motion imposes the constraint
\be
\xi \alpha+ \del_\mu A^\mu = 0 \quad \Rightarrow  \quad \alpha= -\frac{1}{\xi} (\del_\mu A^\mu)\,.
\ee
In the light-cone gauge $A^s= 0$, this relation reads
\be
\alpha =-\frac{1}{\xi} (\del_\rho A^\rho + \del_I A^I)\,,
\ee
which, for the value $\xi= -1$, exactly reproduces the light-cone constraint equation~\eqref{constraint-alpha}.

%******************************************************************************************************************

\end{document}